\documentstyle[preprint,aps]{revtex}
\newcommand{\be}{\begin{equation}} \newcommand{\ee}{\end{equation}}
\newcommand{\bea}{\begin{eqnarray}}\newcommand{\eea}{\end{eqnarray}}

\begin{document}
\draft
\preprint{IP/BBSR/93-62}
\title { Bogomol'nyi Equations of Maxwell-Chern-Simons Vortices from
a generalized Abelian Higgs Model}
\author{Pijush K. Ghosh\cite{mail}}
\address{Institute of Physics, Bhubaneswar-751005, INDIA.}
\maketitle
\begin{abstract}
We consider a generalization of the abelian Higgs model with a
Chern-Simons term by modifying two terms of the usual Lagrangian.
We multiply a dielectric function with the Maxwell kinetic energy
term and incorporate nonminimal interaction by considering
generalized covariant derivative. We show that for a particular
choice of the dielectric function this model admits both
topological as well as nontopological charged vortices satisfying
Bogomol'nyi bound for which the magnetic flux, charge and
angular momentum are not
quantized. However the energy for the topolgical vortices is
quantized and
in each sector these topological vortex solutions are infinitely
degenerate. In the nonrelativistic limit, this model admits
static self-dual soliton solutions with nonzero finite
energy configuration. For the whole class of dielectric function for
which the nontopological vortices exists in the relativistic theory,
the charge density satisfies the same Liouville equation in the
nonrelativistic limit.
\end{abstract}
\narrowtext

\newpage

\section {Introduction}
Recently, the charged vortex solutions\cite{paul86} in the abelian
Higgs model with a Chern-Simons(${\bf CS}$) term have received
considerable attention in the literature because of their possible
relevance in context of cosmic strings as well as planar condensed
matter system. More recently, relativistic model in which the gauge
field is solely given by the ${\bf CS}$ term and the usual $ \mid
\phi \mid^4 $ potential is replaced by $\mid \phi \mid^6$ -type
potential have been considered\cite{hong}.
For a specific choice of the coupling costant, minimum energy
topological\cite{hong,jackiw} as well as nontopological
vortex\cite{jackiw,khare91} configuration arise in
this model that satisfy certain first order differential equation
consistent with the second order equations of motion. These first
order differential equations are known as Bogomol'nyi
equations\cite{bogomol'nyi}. Subsequently,
Lee et. al.\cite{lee252} showed that even the usual $\mid \phi \mid^4$
-type abelin Higgs model (with a specific choice of the coupling
constant) with both Maxwell and the ${\bf CS}$ (${\bf MCS}$) term
admits minimum energy vortex solutions satisfying Bogomol'nyi bound.
However this was possible by the addition of an extra neutral scalar
in the theory. Besides the Gauss Law equation is still second and not
first order in nature.

In the nonrelativistic(${\ NR}$) limit, the pure {\bf CS} theory
provides a second quantized
description of point particles moving(nonrelativistically) in
${\delta}$- function potentials and interacting via {\bf CS}
term\cite{pi}. The matter equation of motion becomes a gauged
nonlinear Schrodinger equation. For static self-dual soliton solution
of this theory, the charge density solves Liouville equation. The
nonrelativistic limit of {\bf MCS} theory also have been discussed
and shown to admit static self-dual soliton solutions\cite{dunne}.
However unlike corresponding pure ${\bf CS}$ case, no analytical
solution can be obtained in this case.
In the nonrelativistic limit, self-dual solitons for both the pure
${\bf CS}$ and ${\bf MCS}$ theory
are of zero energy configuration.

In this paper we consider a generalization of the abelian
Higgs model with a ${\bf CS}$ term, in which we have a
``dielectric function" multiplying the
Maxwell term and an extra gauge invariant nonminimal contribution
to the covariant derivtive. Specifically we are interested in the case
where dielectric
function depends on the Higgs field. We show that for a class of
dielectric function and for a specific choice of the coupling
constant this model admits both topological and nontopological
static minimum energy charged vortex solutions. The Gauss law
equation for these vortex
solutions are of first order in nature and not second order like the
other model for self-dual {\bf MCS} vortices considered by
Lee et. al\cite{lee252}.
Remarkably enough, for a particular
choice of the
dielectric function the first order Bogomol'nyi equations obtained
in this generalized Maxwell-Chern-Simons({\bf GMCS}) theory
can be mapped into the corresponding equations of pure
${\bf CS}$ vortices\cite{hong} upto a scale transformation of the
variables. Also the Bogomol'nyi equations of Torres
model\cite{torres} where an anomolous
magnetic moment contribution plays important role are obtained as
a special case.

The general feature of topological vortices, both
neutral\cite{niel}
and charged, in gauge
theories known to date is that the magnetic flux of the vortices are
necessarily quantized. Furthermore, the vortex solution in each
topolgical sector is nondegenerate. However, we show in this paper
that for a very special
choice of the dielectric function, this model admits topological
charged vortex solutions obeying Bogomol'nyi bound for which
the magnetic flux, and hence the charge and the angular momentum
need not necessarily be quantized even though the energy is quantized.
The topological vortex
solutions are infintely degenerate in energy in each sector and
these degenerate
vortex solutions in a particular sector differ from each other by
flux, charge and the angular momentum. In particular, the topological
vortex solutions in each sector are characterized by energy
$E = {{\pi \kappa^2 n} \over {2 e^2}}$, flux $\Phi = {{2 \pi (n-
\beta)} \over e}$,
angular momentum $J = {{\pi \kappa (\beta^2-n^2)} \over e^2}$ and
charge $Q= -\kappa \Phi$ (where $n$ is the winding number, $\kappa$ is
the coefficient of the {\bf CS} term, $e$ is a coupling constant and
$\beta$ is a parameter describing the solutions).
We derive the sum rules\cite{khare277} for these topological vortices
using which we find that $\beta$
is restricted as, ${1 \over 4} < \beta <n$.

We also study the nontopological vortex solutions present in this
model for the whole class of the dielectric function. The energy,
magnetic flux, charge and angular momentum of these nontopological
vortices are not quantized and determined in terms of a
constant which is the assymptotic value of the gauge field ( with
our choice of the ansatz ) at large distances. We also derive sum
rules for these nontopological vortices. Using these sum rules we show
that lower bound on the energy, magnetic flux,charge and angular
momentum can be obtained for some particular choices of the dielectric
function.

Furthermore, we study the ${NR}$ limit of our model and obtain
static self-dual soliton solutions. However, self-dual soliton
solutions in the nonrelativistic limit is possible only when a
particular relation between two paramaters ( one characterizing the
dielectric function and the other characterizing
the scalar potential), to be discussed later, is satisfied.
This particular constraint, in order to have nonrelativistic self-dual
soliton solutions, is not present in the relativistic theory.
The interesting feature of these
${NR}$ soliton solutions is that unlike corresponding pure
${\bf CS}$\cite{pi}
and ${\bf MCS}$\cite{dunne} case they saturate the lower bound at
some finite nonzero value of the static energy functional. For the
whole class of the dielectric function for which nontopological vortex
solutions is
possible in the relativistic theory, the charge density in the
nonrelativistic theory satisfies the same Liouville equation which is
completely integrable.

We organize the paper as following. In subsec. II.A we set up the
the relativistic model describing in detail the motivation behind
considering such a Lagrangian.
We also obtain second order equations of motion and the
expression for the energy functional. The Bogomol'nyi equations are
obtained in subsec. II.B. In subsec. II.C we study both the
topological and nontopological vortex solutions alongwith their
physical properties. We also show that our model admits a new type
of topological vortex solutions
for which the magnetic flux, charge and the angular momentum is not
quantized and these solutions are infinitely degenerate in each
sector. In subsec. II.D we rigorously show that a bound can be
put on magnetic flux, charge and angular momentum
for both nontopological as well as topological vortices by deriving
sum rules. We take nonrelativistic limit of the relativistic model
and set up the equations of motion as well as the
energy functional in subsec. III.A. In subsec. III.B we obtain static
self-dual soliton solutions in the nonrelativistic model. Finally we
conclude describing our results in brief and indicating future
directions in sec.IV.

\section { Relativistic Theory }
\subsection { The Model }

\noindent We first define our theory by writing the Lagrangian density
\be
{\cal L} \ = \ - \ {\ 1 \over 4} G({\mid \phi \mid})  F_{\mu \nu}
F^{\mu \nu} + \ {1 \over 2} D_{\mu} {\phi} (D^{\mu}{\phi})^* \
+ \ {\kappa \over 4}  \epsilon^{\mu \nu \lambda}  A_{\mu}  F_{\nu
\lambda} - \ V({\mid \phi \mid})
\label{eq:rl}
\ee
where $G({\mid \phi \mid})$ is the scalar field dependent
dielectric fuction and
the generalized covariant derivative is given by
\be
 D_\mu \phi \ = \ (\partial_{\mu}  \ - \ i e A_\mu
 - \ {{\ i g} \over 4} G(\mid \phi \mid) \epsilon_{\mu \nu \lambda}
F^{\nu \lambda})\phi
\label{eq:cd1}
\ee

Our notation is $F_{\mu \nu}={\partial_\mu} {A_\nu} -{\partial_\nu}
{A_\mu}$, $\mu =(0,1,2)$, $g_{\mu \nu}=diag (1,-1,-1)$. We
choose $c= \hbar =1$ throughout this paper
except in section III where we discuss the ${\bf NR}$ limit of this
model and keep $c$ explicitly.

 With a choice of symmetry breaking potential and using the
covariant derivative (\ref{eq:cd1}), both  mass term for the gauge
field and the ${\bf CS}$ term can be generated via spontaneous
symmetry breking({\bf SSB}) mechanism\cite{paul87}.
In fact the nonminimal part of the covariant derivative generates
the {\bf CS} term after {\bf SSB}. Also it is interesting to note
that for $G(\mid \phi \mid)$=1
the nonminimal part can be interpreted as the anomoulous magnetic
moment\cite{kogan}. This is due to the fact that in 2+1 spac-time
Dirac matrices obey
$SO(2,1)$ algebra and Pauli coupling can be incorporated in the
generalized covariant derivative even for the scalar field without
any reference to the spin degrees of freedom.

 The modification to the Maxwell kinetic term can be can be viewed
as an effective action for a system in a medium described by a
suitable dielectric function. In fact, soliton bag
models\cite{friedberg} of quarks and gluons
are described by Lagrangian, where such a dielectric function is
multiplied with the Maxwell kinetic energy term. Also in certain
supersymmetric theories with a non-compact gauge group\cite{hull},
such a nonminimal kinetic term was necessary in order to have a
sensible gauge theory. In the context of vortex solutions, this
non-minimal coupling is interesting because of the existence of
Bogomol'nyi bounds for a more general form of the scalar
potential\cite{nam,lohe}. Lee et. al.\cite{nam} considered
the Lagrangian (\ref{eq:rl}) with ${\kappa=0}$ and without the
nonminimal contribution to the covariant derivative (\ref{eq:cd1})
and have shown that the
model admits topological as well as nontopological static
self-dual neutral vortex solutions. Torres\cite{torres} considered
the Lagrangian (\ref{eq:rl}) with ${\ G(\mid \phi \mid)}=1$ and
obtained static minimum energy nontopological vortex configuration
for a simple ${\mid \phi \mid}^2$ potential. As a natural extension,
we consider the effect of both the dielectric function and the
generalized covariant derivative for arbitrary ${\ G(\mid \phi \mid)}$
and study the Bogomol'nyi
limit for topological as well as nontopological vortex solution.

 The equations of motion for the Lagrangian (\ref{eq:rl}) are
\be
D_{\mu}D^{\mu}\phi \ = \ - \ 2 {{\partial V(\mid \phi \mid)} \over
{\partial \phi^*}} \ - {\ 1 \over 2} {{\partial G(\mid \phi \mid)}
\over {\partial \phi^*}} F_{\mu \nu} F^{\mu \nu}
 \ - \ {\ g \over 2 e} {{\partial G(\mid \phi \mid)} \over
{\partial \phi^*}} \epsilon^{\mu \nu \lambda} J_\mu F_{\nu \lambda}
\label{eq:sfr}
\ee
\be
\epsilon_{\mu \nu \lambda} \ {\partial^\mu} \ [G(\mid \phi \mid) \
(F^\lambda \ + {\ g \over 2 e} J^\lambda)] \ = \ J_\nu \ - \
\kappa F_\nu
\label{eq:gfr}
\ee
where the dual field $F_\mu$ and the conserved current $J_\mu$ is
\be
F_\mu \ = \ {\ 1 \over 2} \epsilon_{\mu \nu \alpha} \  F^{\nu \alpha}
\label{eq:dual}
\ee
\be
J_\mu \ = \ - \ {{\ i e} \over 2} [\phi^* D_\mu \phi \
- \ \phi (D_\mu \phi)^*]
\label{eq:cdr}
\ee

The energy momentum tensor $T_{\mu \nu}$ is obtained by varying the
curved space form of the action with respect to the metric
\bea
T_{\mu \nu} \ &  = & \ G({\mid \phi \mid}) \ [1-
{\ g^2 \over 4} G({\mid \phi \mid}) {\mid \phi \mid}^2] \ [F_\mu F_\nu
- {\ 1 \over 2} g_{\mu \nu} F_\alpha F^\alpha]\nonumber \\
& & + { \ 1 \over 2} [\bigtriangledown_\mu \phi
(\bigtriangledown_\nu \phi)^*+{\bigtriangledown_\nu \phi}
({\bigtriangledown_\mu \phi})^*]\nonumber \\
& & - \ g_{\mu \nu} [{\ 1 \over 2} {\mid {\bigtriangledown_\alpha}
\phi \mid }^2 - V ({\mid \phi \mid}) ]
\label{eq:emtr}
\eea

\noindent where $\bigtriangledown_\alpha \ = \ \partial_\alpha \ - \
i e A_\alpha$ includes
only the gauge potential contribution.

\subsection {Bogomoln'nyi Equation}

\noindent Now notice that the first order equation
\be
J_\nu \ = \ k F_\nu
\label{eq:fgf}
\ee
solves the gauge field equations (\ref{eq:gfr}) for
arbitrary $G({\mid \phi \mid})$
provided the following relation among the coupling constant holds
\be
g \ = \ - \ {\ 2e \over \kappa}
\label{eq:cc}
\ee
\noindent Like pure {\bf CS} Higgs theory\cite{hong}, the zero
component of (\ref{eq:fgf}), i.e, Gauss law implies that the solution
with charge $Q$ also carries
magnetic flux ${\Phi} \ = \ - \ {Q \over \kappa}$. It should be noted
that for ${G({\mid \phi \mid}) \neq 0}$ equation (\ref{eq:fgf}) is
essentially different from
that of corresponding equation for pure ${\bf CS}$ vortices as $D_\mu$
receives contribution from the nonminimal part also. In particular,
using equations (\ref{eq:cdr}) and (\ref{eq:cc}) the gauge field
equation (\ref{eq:fgf}) can be
rewriten as
\be
\kappa \ F_\mu =
J_\mu \ = \ (1 - {e^2 \over \kappa^2} G(\mid \phi \mid) {\mid \phi
\mid}^2)^{-1} \tilde{J_\mu}
\label{eq:nfgf}
\ee
\noindent where $\tilde{J_\mu}$ receives only minimal contribution
\be
\tilde{J_\mu} \ = \ - \ {{\ i e} \over 2} [\phi^* \bigtriangledown_
\mu \phi \ - \ \phi (\bigtriangledown_\mu \phi)^*]
\label{eq:ncdr}
\ee
\noindent We notice that for the choice of $G(\mid \phi \mid) =
{\kappa^2 \over e^2} (1-C_{0}) {\mid \phi \mid}^{- 2}$,
equations (\ref{eq:nfgf}) is also the gauge field equation for the
pure ${\bf CS}$ Higgs theory except the overall constant
factor $1/C_{0}$ multiplying with $\tilde{J_\mu}$. This implies that
the electric and magnetic field comonents of pure {\bf CS} Higgs
theory and our model with the above choice
of the dielectric function differ by the
scale factor $1/C_{0}$. For $C_{0}=1$,i.e, $G=0$,
our model reduces to the pure {\bf CS} Higgs theory as can be seen
from (\ref{eq:rl}) and (\ref{eq:cd1}). It may therfore be worthwhile
to consider the
possibility of obtaining Bogomol'nyi limit, for the choice
$G(\mid \phi \mid) = {\kappa^2 \over e^2}
(1-C_{0}) {\mid \phi \mid}^{- 2}$. However we are interested in a
more general class of dielectric function and the scalar potential.
So we keep $G(\mid \phi \mid)$ arbitrary unless mentioned otherwise
and obtain the results
for above mentioned choice of $G(\mid \phi \mid)$ as a special case.
Using equation (\ref{eq:fgf}) and (\ref{eq:cc}), we rewrite
the scalar field equation of motion (\ref{eq:sfr})
\be
D_{\mu} D^{\mu} \phi \ = \ - \ 2 {{\partial V(\mid \phi \mid)} \over
{\partial \phi^*}} \ + \
{\ 1 \over 2} {{\partial G(\mid \phi \mid)} \over {\partial \phi^*}}
F_{\mu \nu} F^{\mu \nu}
\label{eq:sfr2}
\ee

 Now we seek vortex solutions in the system described by
the equations (\ref{eq:nfgf}) and (\ref{eq:sfr2}),
when the relation (\ref{eq:cc}) is satisfied.
We choose the ansatz for rotationally symmetric solution of winding
number $n$
\be
\vec{A(r)} \ = \ - \ {\hat \theta} {{a(r) \ - \ n} \over {e r}} ,
A_0 (r) \ = \ {\ k \over e} h(r) ,
\phi(r) \ = \ {\ k \over e} f(r) e^{- i n \theta}
\label{eq:ansatz}
\ee
\noindent After substituting the ansatz (\ref{eq:ansatz}),
equations (\ref{eq:nfgf}) and (\ref{eq:sfr2}) can be reduced to
\be
{\ 1 \over r} \ [1 \ - \ G(f) f^2] a^\prime \ + \ k^2 f^2 h
\ = \ 0
\label{eq:rs1}
\ee
\be
r [1 \ - \ G(f) f^2 ] h^\prime \ + \ a f^2 \ = \ 0
\label{eq:rs2}
\ee
\be
{\ 1 \over r} {{\partial} \over {\partial r}} (r {{\partial f} \over
{\partial r}}) \  +  {\ k^2 f \over {\ ( 1\ - \ G(f) f^2 )^2}} (1 \
+ \ {\ f^3 \over 2} {{\partial G(f)} \over {\partial f}}) (h^2 \ -
{\ a^2 \over {\ k^2 r^2}})  =  \ {\ e^2 \over k^2} {{\partial V}
\over {\partial f}}
\label{eq:rs3}
\ee
\noindent where prime denotes differentition with respect to r.
The energy functional that is obtained
from equation (\ref{eq:emtr}) for the ansatz (\ref{eq:ansatz}) is
\bea
E \ & = & \ {\ k^2 \over 2 e^2} {\cal \int} d^2 x \{ G(f) (1 \ -
G(f) f^2) [ (h^\prime)^2 \ + \
({\ a^\prime \over k r})^2 ] \ + \ (k h f)^2\nonumber \\
& & + \ (f^\prime)^2 \ + \
({f a \over r})^2 \ + \
{\ 2 e^2 \over k^2} V(f) \}
\label{eq:rse}
\eea
After eliminating $h(r)$ and $h^\prime(r)$ from equation
(\ref{eq:rse}) using equations
(\ref{eq:rs1}) and (\ref{eq:rs2}) the energy functional can be
written solely in terms of $f(r)$ and $a(r)$
\be
E={\ k^2 \over 2 e^2} {\cal \int} d^2 x [ (f^\prime)^2+{\ {f^2 a^2}
\over {\ r^2 (1-G(f) f^2)}}+({\ a^\prime \over {\ kr}})^2 {\ (1-
G(f) f^2) \over \ f^2}+{\ 2 e^2 \over k^2} V(f) ]
\label{eq:rse2}
\ee

Let us a define a function $W(f)$ shortly to be fixed in terms of
$G(f)$. The energy functional (\ref{eq:rse2}) can be rearranged
using Bogomol'nyi trick as,
\bea
E & = & {\kappa^2 \over {2 e^2}} \int d^2 x \{ [f^\prime \pm {{f a}
\over {r (1- G(f) f^2 )^{1 \over 2}}}]^2 + {{1- G(f) f^2} \over f^2}
[{a^\prime \over {\kappa r}} \mp W]^2 +{{2 e^2} \over \kappa^2}
V(f)\nonumber \\
& & - {{(1-G(f) f^2) W^2} \over f^2} \}
\pm {{2 \pi \kappa^2} \over {\sqrt{C_{0}} (1+\gamma)e^2}} [R(\infty)
-R(0)]
\label{eq:rse3}
\eea
\noindent where
\be
R(r)  =  {{\sqrt{C_{0}} (1+\gamma)} \over {\kappa f^2}}
(1- G(f) f^2)^{{\gamma+1} \over 2} a W
\label{eq:rse4}
\ee
\noindent and $W(f)$ is determined by
\be
{d \over {d r}} [ {{1 - G(f) f^2} \over {\kappa f^2}} W ]
= -  \ {{f^\prime f} \over {(1-G(f) f^2)^{1 \over 2}}}
\label{eq:rse5}
\ee
\noindent However for arbitrary $G(f)$ no analytical solution
can be obtained for $W(f)$. For simplicity, we choose $G(f)$ to be
\be
G(f)=f^{\ -2} - C_{0} \ f^{- 2} \ (1-f^2)^{\ 1- \gamma}
\label{eq:df}
\ee
\noindent where $\gamma$ is a real number and $C_{0}$ is a
positive constant. The function $W(f)$ for this choice of $G(f)$
follows from equation (\ref{eq:rse5})
\be
W(f) = {{\kappa f^2} \over {C_{0}^{3 \over 2} (1+\gamma)}} (1-
f^2)^{{3 \gamma - 1} \over 2}
\ee
For this choice of $G(f)$, and hence $W(f)$, the energy functional
can be written as
\bea
E \ &  = & \ {\ k^2 \over {\ 2 e^2}} {\cal \int} d^2 x \{ [f^\prime
\ \pm \ {{\ f a} \over {\ {C_{0}}^{1 \over 2} r}} (1-
f^2)^{{\gamma \ - \ 1} \over
2}]^2 \nonumber \\
& & + C_{0} f^{\ -2} (1-f^2)^{\ 1- \gamma} \ [{{\ a^\prime} \over
{\kappa  r}} \ \mp \
{{\kappa} \over {\ C_{ 0}^{\ 3 \over 2} (1+\gamma)}} f^2 (1-
 f^2)^{{\ 3 \gamma - 1} \over 2}]^2 \nonumber \\
& & +{\ 2 e^2 \over k^2} V(f)-
{{\kappa^2} \over {\ C_{ 0}^2 (1+\gamma)^2}}
f^2 (1-f^2)^{\ 2 \gamma}\} \ \pm \ {{2 \pi \kappa^2} \over {\ C_{ 0}
^{ 1 \over 2} (1+\gamma) e^2}} [ R(\infty)-R(0) ]
\label{eq:be}
\eea
\noindent where
\be
R(r) = a (1-f^2)^{{1+\gamma} \over 2}
\label{eq:be0}
\ee
\noindent When $\gamma$ is odd integer
there
is a lower bound on the
energy provided we choose the scalar potential
\be
V(f)={{\kappa^4} \over {\ 2 C_{0}^2 e^2 (1+\gamma)^2}} f^2 (1
-f^2)^{\ 2 \gamma}
\label{eq:sp}
\ee
\noindent The lower bound on
the energy also exists for arbitrary $\gamma$ ( and above choice of
$V(f)$ ) provided $0 \leq f \leq 1$. The lower
bound on energy is saturated when the following Bogomol'nyi equations
are satisfied
\be
f^\prime = \pm
\  \ {{\ f a} \over {\ {C_{0}}^{1 \over 2} r}} (1-
f^2)^{{\gamma - 1} \over
2}
\label{eq:fbe1}
\ee
\be
{{\ a^\prime} \over
{\kappa  r}} \ = \ \mp  \
{{\kappa} \over {\ C_{0}^{\ 3 \over 2} (1+\gamma)}} f^2 (1-
 f^2)^{{\ 3 \gamma - 1} \over 2}
\label{eq:fbe2}
\ee
\noindent One can easily check that these two first order differential
equations are consistent with the second order differential equation
(\ref{eq:rs3}). At the Bogomol'nyi limit, two diagonal elements of
the energy momentum tensor other than $T_{\ 0 0}$, i.e,
$T_{\ r r}$ and $T_{\theta \theta}$ vanishes. The off-diagonal
elment $T_{\ t \theta}$ is however nonvanishing
\be
T_{\ t \theta}=- {\kappa \over {\ e^2 r^2}} a^\prime a
\label{eq:am1}
\ee
implying that the solution to the Bogomol'nyi equations carry finite
angular momentum for well behaved $a(r)$
\be
J={{\pi \kappa} \over e^2} \ [a(\infty)^2-a(0)^2)]
\label{eq:am2}
\ee
\noindent Note that the following scale transformation
\be
a \rightarrow C_{0}^{\ 1 \over 2} \ a, \  r \rightarrow C_{0} \ r,
f \rightarrow \ f
\label{eq:st}
\ee
\noindent eliminates $ C_{ 0}$ from the equations (\ref{eq:fbe1})
and (\ref{eq:fbe2}). The
decoupled second order equation for the $f(r)$ is
\be
f^{\prime \prime}+{\ f^\prime \over r}+
{\kappa^2 \over {\ \gamma+1}} f^3 (1-f^2)^{\ 2 \gamma -1}-
{{\ f^\prime}^2 \over f}+({\gamma-1}) {\ {{f^\prime}^2 f} \over
{\ 1-f^2}}=0
\label{eq:ds}
\ee
\noindent where the scale transformation (\ref{eq:st}) have been
performed. Equation (\ref{eq:ds}) is highly nonlinear and we do
not have the analytical solution for
it. However we can obtain asymptotic as well as numerical solution
for this equation. We do so while discusing vortex solution for
different choices of $\gamma$.

\subsection {Vortex Solution}

In this section we discuss assymtotic vortex solutions for well
behaved scalar
potential. We choose either $\gamma =  0$ or $\gamma$ odd
so that the scalar potential
is bounded from below and $f(r)$ is not restricted.

I. \ \ $\gamma = 0 \ $, i.e, $G=(1-C_{0}+C_{0} f^2) f^{\ -2}$ \ :
 The Bogomol'nyi
bound arises when the scalar field mass $m$ and the {\bf CS}
coefficient $\kappa$ are related as, $m={\kappa \over C_{0}}$.
For $C_{0 }=1$, i.e, $G=1$; our model reduces to that of
considered by Torres\cite{torres}. For arbitrary $C_{0}(C_{0}>0)$,
Bogomol'nyi equations of
our model can be mapped into corresponding equations of Torres model
upto a scale transformation of the variables. For this case only
nontopological vortices exists. The flux, charge and angular momentum
of these nontopological vortices are independent of $C_{0}$ and same
for both Torres model as well as our model. However, the energy
$E = {\kappa^2 \over {\sqrt{C_{0}} e}} \Phi$ is parametrized by
$C_{0}$ and distinguishes between Torres model and our model.

II. \ \ $\gamma=1$, i.e, $G=(1-C_{0}) f^{\ -2}$ :
For $C_{0}=1$, i.e, $G=0$, our model
reduces to pure ${\bf CS}$ Higgs theory. Equations
(\ref{eq:sp}) to (\ref{eq:fbe2})
reproduce the Bogomol'nyi equations as well as the potential of
pure ${\bf CS}$ Higgs theory. Even for arbitray $C_{0}(C_{0}>0)$
when Maxwell term is present
, these Bogomol'nyi equations of {\bf GMCS} are exactly same
as those of corresponding
pure ${\bf CS}$ theory\cite{hong} provided the
scale transformation (\ref{eq:st}) is performed. This implies that
the Bogomol'nyi equations of ${\bf GMCS}$ Higgs theory
can be mapped into the
Bogomol'nyi equations of pure ${\bf CS}$ Higgs theory upto a scale
transformation of the variables. Hence both topological as well as
nontopological vortex solutions also exists in this case. The
topological vortex solutions are characterized by the flux
$\Phi^{top.}={{2 \pi} \over e} n$, the energy $E^{top.}=
{\kappa^2 \over {\sqrt{C_{0}} e}} \Phi^{top.}$,
the charge $Q^{top.}=- \kappa \Phi^{top.}$ and the angular momentum
$J^{top.}= - {{\pi \kappa} \over e^2} n^2$. The nontopological
vortex solutions are characterized by the flux $\Phi^{nontop.}=
{{2 \pi} \over e} (n+\alpha)$, the energy $E^{nontop.}= {\kappa^2
\over {\sqrt{C_{0}} e}} \Phi^{nontop.}$, the charge $Q^{nontop.}=-
\kappa \Phi^{nontop.}$ and the angular momentum
$J^{nontop.}= {{\pi \kappa} \over e^2} (\alpha^2-n^2)$, where $\alpha$
is a parameter describing the nontopolgical solutions. We shall see in
subsec. II.D that $\alpha \geq n+2$.

III. \ \ $\gamma > 1 $: \ The finiteness of the energy can be ensured
by requireing either (i)$a(\infty)= \ - \alpha$, $f(\infty)=0$ or
(ii)$a(\infty)={\beta}$, $f(\infty)=1$. Here $\alpha$
and ${\beta}$ are two real positive constants.
Further, on demanding nonsingular field variables, boundary condition
at the origin gets fixed as (iii)$ a(0)=n \ , \  f(0)=0$ for $n\neq0$,
(iv)$a(0) = 0 \ , \ f(0) = G_{0}$ for $n=0$.
Since the solutions for $n$ and $-n$ are related by the transformation
$f \rightarrow f$, $a \rightarrow -a$; we consider only the case
$n \geq 0$, further without any loss of generality we choose
$C_{0}=1$. Notice that for $\gamma > 1$, the constant $C_{0}$ does
not play any special role. The boundary condition (i) corresponds
to nontopological vortex solution
while the boundary condition (ii) corresponds to topological vortex
solution which we now discuss in some detail.

(a) {\bf Nontopological vortex solutions} : For nontopological
vortices, at large distances $f(r) \rightarrow 0$ and
$a(r) \rightarrow - \alpha$. The power series solution of equations
(\ref{eq:fbe1}), (\ref{eq:fbe2}), (\ref{eq:ds}) can be
shown to be
\be
f(r)= \ {\ A \over (\kappa r)^{\alpha}} \ - \ {\ A^3 \over
{\ 4 (\gamma+1) (\alpha-1)^2 (\kappa r)^{\ 3 \alpha-2}}} \ + \
{\bf O} ( ({1 \over {\kappa r}})^{3 \alpha} )
\label{eq:ai1}
\ee
\bea
a(r) \ & = &  \ -\alpha \ + \ {\ A^2 \over {2 (\gamma+1) (\alpha-1)
(\kappa r)^{\ 2 \alpha-2}}}\nonumber \\
& & - \ {\ A^4 \over {\ 8 (\gamma+1)^2
(\alpha-1)^3 (\kappa r)^{\ 4 \alpha-4}}} \ + \
{\bf O} ( ({1 \over {\kappa r}})^{\ 4 \alpha-2})
\label{eq:ai2}
\eea
\noindent The behavior at small distances when $n \neq 0$
is given by
\be
f(r) = \ B (\kappa r)^n \ - \ { \ {\gamma - 1} \over 4}
B^3 (\kappa r)^{ 3 n} \ + \
{\bf O}((\kappa r)^{ 3 n+2} )
\label{eq:ai3}
\ee
\bea
a(r)  = n  - \ {{ B^2  (\kappa r)^{\ 2 n+2}} \over {2 (\gamma+1)
(n+1)}} \ + \ { {(2 \gamma-1 ) B^4 (\kappa r)^{ 4 n+2}}  \over {2 (2
n \ + \ 1) (\gamma+1)}}\ + \ {\bf O}((\kappa r)^{6 n+2})
\label{eq:ai4}
\eea
These solution are characterized by magnetic flux $\Phi={{2 \pi (n+
\alpha)} \over e}$, charge $Q= - \kappa \Phi$,
energy $E = {{\kappa^2} \over { (1+\gamma) e}} \Phi$
and angular momentum $J = {{\pi \kappa (\alpha^2-n^2)}\over e^2}$.

 As in Ref. \cite{jackiw}, the constant $B$ is not determined by the
behaviour of the fields
near the origin, but is instead fixed by requiring proper behaviour
as $r \rightarrow \infty$. In particular, $B$ is a function of the
assymptotic value of the gauge field, i.e, $\alpha$. Now notice from
equation (\ref{eq:ai2}) that $\alpha > 1$. Since the second term in
(\ref{eq:ai2}) is subleading compare to the first term only in that
case. So, all values of $B$ are not allowed in order to have
nontopological vortex solutions. It turns out numerically that
for each integer $n$ a continuous set of nontopological vortex
solutions exists corresponding
to the range $0 < B < B_{0} (B_{0} < 1)$, where $B_{0}$ is a critical
value of $B$ separating nontopological vortices from topological
vortices. As an illustration, we plot nontopological vortices of this
kind for $\gamma=3$ when $n=1$ and $n=2$ in Fig. 1 and Fig. 2
respectively. From both the figures we observe that $max(f) < 1$.
In fact, the peak of
$f(r)$ gradually decreases as $B$ takes comparatively lower value than
$B_{0}$. We have checked for other values of $B$ close to $B_{0}$ that
$max(f) < 1$. In subsec. II. D, we shall derive sum rules for these
nontopological vortices and using this fact, shall show that
$\alpha \geq n+2$ (for $C_{0}=1$).

For zero vorticity, i.e, $n=0$, $f(0)$
is not constrained. So the power series solution for both $f(r)$ and
$a(r)$ near the origin is quite different
from the $n \neq 0$ case. We find
\bea
f(r) \ & = & \ G_0 \ - \ {{G_0^3 (1-G_0^2)^{2 \gamma-1}} \over
{4 (\gamma+1)}} (\kappa r)^2\nonumber \\
& & + \ {{G_0^5 (1-G_0^2)^{4 \gamma-3}} \over {64 (\gamma+1)^2}} (4-
G_0^2-5 \gamma G_0^2) \ (\kappa r)^4 \ + \ {\bf O} ( (\kappa r^6) )
\label{eq:ai5}
\eea
\bea
a(r) \ & = & \ - {{G_0^2 (1-G_0^2)^{{3 \gamma-1}
\over 2}} \over {2 (\gamma+1)}} (\kappa r)^2\nonumber \\
& & + \ {{G_0^4 (1-G_0^2)^{{7 \gamma-5} \over 2}}
\over {16 (\gamma+1)^2}} (2 \ - \ G0^2 \ -
\ 3 \gamma G_0^2) \ (\kappa r)^4 +
{\bf O} ( (\kappa r)^6)
\label{eq:ai6}
\eea
\noindent We plot a nontopological soliton of this kind for
$\gamma=3$ in Fig. 3.

(b) {\bf Topological vortex solutions} : For simplicity, we
restrict our discussion on the topological vortex solutions
to $\gamma=3$.
The behaviour of the field
variables near the origin for $n \neq 0$ can be obtained from
(\ref{eq:ai3}) and (\ref{eq:ai4}) by putting $\gamma=3$.
We obtain large distance behaviour as following
\be
f(r) \ = \ 1 + {D \over (\kappa r)^{2 {\beta}}} +
{{3 D^2} \over {2 (\kappa r)^{4 {\beta}}}} \ +
\ {\bf O} ( \ ({1 \over {\kappa r}})^{6 \beta})
\label{eq:ai7}
\ee
\be
a(r) \ = \ {\beta} + {{2 D^4} \over {(4 {\beta}-
1) (\kappa r)^{8 {\beta}-2}}} + {{16 D^5 } \over {(5 {\beta} \ - \
1) (\kappa r)^{10 {\beta}-2}}} \
+ \ {\bf O} ( \ ({1 \over \kappa r})^{12 \beta \ - \ 2})
\label{eq:ai8}
\ee
\noindent Note that the large distance behaviour of the scalar
field and the gauge field for these topological vortices are of
semi-local\cite{semi-local} type, i.e, they fall off obeying
power law.

 It is remarkable to note that when
the scalar field $f(r)$ attains its asymmetric vaccum value at large
distances, $a(r)$ does not vanish; a feature not known so far for the
topological vortices. The novel consequnce is that even for
topological vortices the magnetic flux, and hence the charge and
the angular momentum,
need not necessarily be quantized; while the energy, as evident
from (\ref{eq:be}), is quantized. In particular, these topological
vortex solutions are characterized by energy $E={{\pi \kappa^2 n}
\over {2 e^2}}$, flux $\Phi={{2 \pi} \over e} (n-\beta)$, charge
$Q= - \kappa \Phi$ and angular momentum $J={{\pi \kappa}
\over e^2} (\beta^2-n^2)$. Since for each $n$, there is a set
of solutions paramterized by $\beta$; the solutions are degenerate
in each topological sector
and one solution differs from another by charge, flux and angular
momentum. At this point its worthwhile to ask whether $\beta$ can
take any positive value or is bounded from above and/or below. We
find that $\beta$ is indeed bounded from above and below, i.e,
${1 \over 4} <\beta < n$. The lower bound on $\beta$ is due to the
fact that the second term in (\ref{eq:ai8}) is subleading compare
to the first term only when $\beta > {1 \over 4}$. The upper bound
follwos from the sum rules for these topological vortices,
derived below in subsec. II.D.

As in nontopological vortex solutions, $B$ is a function of $\beta$.
Since ${1 \over 4} < \beta < n$, hence $B$ is also restricted. We find
numerically that for each $n$ a continuous set of topological vortex
solutions exists corresponding to the range $B_{0} < B <1$.
We plot $f(r)$ (solid line) and $a(r)$ (dashed line)
in Fig. 4 for (I) $n=1, \beta=0.78$;
(II) $n=2, \beta=1.91$; (III) $n=2, \beta=1.53$ and (IV) $n=2,
\beta=1.23$. We have given only one plot for $n=1$ since the profile
of $f(r)$ and $a(r)$ are almost same for different $\beta$.

 The Bogomol'nyi equations (\ref{eq:fbe1})
and (\ref{eq:fbe2}) are also
satisfied by the trivial vacuum solution $f(r)=1$, $a(r)=n$. This
particular topological vortex solution has zero energy, charge, flux
and angular momentum. Usually any model admitting topological vorex
solutions has trivial vaccum solution only for $n=0$. However in
our model trivial vaccum solution exists in each topological
sector, i.e, for any $n$. We have also looked for a nontrivial
solution with the boundary condition
$a(0) = n, \ f(0) = 1$ for $n \neq 0$. However no well behaved
power series solution near origin is possible for this choice of
boundary condition.

\subsection{Sum Rules}
The magnetic flux, and hence the charge and the angular momentum
of both topological as well as nontopological vortices are determined
in terms of unknown constants which are the assymptotic values of
the gauge field
(with our choice of ansatz) at large distances. We now show that a
bound on these constants can be obtained
rigorously using the first order Bogomol'nyi equations. Infact,
Khare\cite{khare277}
obtained countable infinite number of sum rules for pure ${\bf CS}$
topological as well as nontopological
vortices and using the first two he was able
to put lower bound on the fluxe of nontoplogical vortices. In
the same spirit,
we derive and study the first few
sum rules for the topological as well as nontopological
${\bf GMCS}$ vortices.

The usual technique for deriving sum rules is following.
Let us consider the identity
\be
{1 \over {l+1}} {d \over {d r}} a^{l+1} = a^l a^\prime
\label{eq:identity}
\ee
\noindent where $l$ is any nonnegative integer, i.e, $l=0$,$1$,$2$,...
On integrating both sides with respect to $r$ from $0$ to $\infty$
we find that the left hand side of (\ref{eq:identity}) gives
${1 \over {l+1}} [ a(\infty)^{l+1} -a(0)^{l+1} ]$. The right hand side
of (\ref{eq:identity}) can be simplified step by step by making
use of the Bogomol'nyi equations (\ref{eq:fbe1}), (\ref{eq:fbe2}) and
boundary conditions (i), (ii), (iii). Since the boundary conditions
for the topological vortices are different from those of the
nontopological vortices,
we dicuss these two cases seperately.

(a) {\bf Sum rules for nontopological vortices} : The first sum rule
(i.e, for $l=0$ ) is obtained after integrating equation
(\ref{eq:fbe2}) once with respect to $r$ and using boundary
conditions (i) and (iii)
\be
\alpha \ + \ n \  =  \ {\kappa^2 \over {(1+\gamma) C_{0}^{3 \over 2}}}
{\int_{0}^{\infty}}
r dr f^2 (1-f^2)^{{\ 3 \gamma-1} \over 2}
\label{eq:sr1}
\ee
\noindent In order to obtain the second sum rule (i.e, for $l=1$),
first
we rewrite $a a^\prime$ solely in terms of $f$, $f^\prime$ and $r$,
by using equations (\ref{eq:fbe1}) and (\ref{eq:fbe2}). Then, on
expanding $(1-f^2)^\gamma$ binomially and after integrating the
right hand side of equation (\ref{eq:identity}) by parts and using
the boundary conditions for the nontopological vortices; we have
\be
\alpha^2 \ - \ n^2 \  =  \ {{2 \kappa^2} \over {(1+\gamma)
\sqrt{C_{0}}}} {\int_{0}^{\infty}}
r dr \sum_{p=0}^{\gamma} {\ (-1)^p \over {p+1}} \ \
^\gamma C_p f^{\ 2 (p+1)}
\label{eq:sr2}
\ee
where $^\gamma {C}_p=\gamma !/(p! (\gamma-p)!)$.
\noindent For $\gamma=0$ the second sum rule (\ref{eq:sr2}) is simply
$ \alpha^2 - n^2 = {{2 \kappa^2} \over C_{0}}
{\int_{\ 0}^{\infty}} r dr f^2 > 0$. This implies
that $\alpha \ > \ n$ for arbitrary $C_{0}$ since right hand side
is manifestly positive
definite. Not surprisingly the numerical
calculation\cite{torres} for $C_{0}=1$ is in agreement with this exact
result. Simillarly for $\gamma =1$,
using both the sum rules (\ref{eq:sr1}) and (\ref{eq:sr2}) we find
that $\alpha \geq n+2 \sqrt{C_{0}}$ as in pure
${\bf CS}$ case (i.e, $C_{0}=1$); but now for arbitrary
$ C_{0} (C_{0} > 0 )$, i.e,
for ${\bf GMCS}$ vortices. The magnetic moment of these
nontopological vortices\cite{khare91,khare277}
\be
\mu_{z} = {1 \over 2} \int d^2 r (\vec{r} \times \vec{J})_{z}= -
{{\pi \kappa^2} \over e} \int r^2 dr h^\prime(r)
\label{eq:mm}
\ee
can also
be calculated for $\gamma=1$ using the sum rules (\ref{eq:sr1}) and
(\ref{eq:sr2}). In particular, we find
\be
\mu_{z}^{nontopo.} = - {{2 \pi \kappa^2 \sqrt{C_{0}}} \over e} (\alpha+
n) (\alpha-n-\sqrt{C_{0}})
\label{eq:mm1}
\ee
\noindent Note that $\mu_{z}^{nontopo.}$ is always negative.

For $\gamma >1$, the bound
${\alpha >1}$
which follows from (\ref{eq:ai2}) can not be improved using these two
sum rules alone. This can be seen as follows. Since we
are considering only odd $\gamma$,
the term with highest power in $f(r)$ in the second sum rule
(\ref{eq:sr2}) is
$ - {{2 \kappa^2} \over C_{0}} (\gamma+1)^{-2} f^{2 (\gamma+1)}$,
while for the first sum rule
(\ref{eq:sr1})
it is $\mp {\kappa^2 \over C_{0}^{3 \over 2}} (1+\gamma)^{-1}
 f^{3 \gamma+1}$; where negative sign is
meant for $\gamma=4 m+1$ and positive sign for $\gamma=4 m-1$,
$m=1, 2, 3...$ .
It is obvious that no manifestly positive or negative
definite expression for $f(r)$
is possible for $\gamma=4 m+1$, since $3 \gamma+1 \ > \ 2(\gamma+1)$
when $\gamma>1$. Considering the term with second highest power in
$f(r)$ in second sum rule (\ref{eq:sr2}), the possibility
$\gamma=4 m-1$ is also
ruled out. However if the numerical calculation is any guide, we can
put a lower bound on the magnetic flux of the nontopological
vortices at least for $\gamma=3$. For $\gamma=3$,
the two sum rules (\ref{eq:sr1}) and (\ref{eq:sr2}) can be combined
to write the following expression
\be
\alpha^2 - n^2 - 2 \sqrt{C_{0}} (\alpha + n) =
{\kappa^2 \over {2 C_{0}}} \int_{0}^{\infty} r dr
[ {5 \over 2} f^4 (1-f^2)^2 + f^8 ({5 \over 4} - f^2) ]
\label{eq:sr3}
\ee
\noindent We know from the numerical calculation that $max(f) < 1$
for $n=0$, $1$, $2$
and we believe this is true for $n > 2$ also. So right hand side of
(\ref{eq:sr3}) is manifestly positive definite and
$\alpha \geq n + 2 $ for $C_{0}=1$.

 (b) {\bf Sum rules for topological vortices} : In order to obtain
sum rules for topological vortices we recall that the  boundary
conditions are, $f(\infty)=1$, $a(\infty)={\beta}$, $f(0)=0$,
$a(0)=n$. Following
exactly the same technique we find that the first three sum rules
(i.e, for $l=0$, $l=1$ and $l=2$) are
\be
n - \beta = {\kappa^2 \over {(1+\gamma) C_{0}^{3 \over 2}}}
\int_{0}^{\infty} r dr f^2 ( 1 - f^2 )^{{3 \gamma-1} \over 2}
\label{eq:sr4}
\ee
\be
n^2 - \beta^2 = {{2 \kappa^2} \over {(1+\gamma)^2 C_{0}}}
\int_{0}^{\infty} r dr ( 1 - f^2 )^{1+\gamma}
\label{eq:sr5}
\ee
\bea
n^3 - \beta^3 & = & - {{3 \kappa^2} \over {(1+\gamma)^2 \sqrt{C_{0}}}}
\int r dr \ [ \ {r^2 \over {2 (1+\gamma) C_{0}^{3 \over2}}}
f^2 (1-f^2)^{{5 \gamma+1} \over 2}
+ 2 \log f\nonumber \\
& & - 2 \sum_{p=1}^{p={{\gamma+3} \over 2}} {(-1)^p \over {2 p}}
 \ \ ^{{\gamma+3} \over 2} C_{p} (1-f^{2 p}) \ ]
\label{eq:sr6}
\eea
\noindent where $\gamma=1$, $3$. When $\gamma=1$, the magnetic moment
of these topological vortices can be calculated as in case of
nontopological vortices. Using equations (\ref{eq:mm}),
(\ref{eq:sr4}) and (\ref{eq:sr5}) we find
\be
\mu_{z}^{topo.} = {{2 \pi \kappa^2 \sqrt{C_{0}}} \over e}
n (n+\sqrt{C_{0}})
\label{eq:mm2}
\ee
\noindent Note that $\mu_{z}^{topo.}$ is positive definite and can be
obtained from $\mu_{z}^{nontopo.}$ which is always negative,
by putting $\alpha=0$.

When $\gamma=3$, for both the first two sum rules (\ref{eq:sr4}) and
(\ref{eq:sr5}) the right hand side is positive definite and this
implies that $\beta \leq n$. When $\beta$ saturates
the upper bound, the Bogomol'nyi equations are satisfied by the
trivial vaccum solution $f(r)=1$, $a(r)=n$ in each topological sector.
We have
already seen that $\beta > {1 \over 4}$ from equation (\ref{eq:ai2}).
So, the infinitely degenerate vortex solutions exists
in each topological sector for ${1 \over 4} < \beta < n$,
characterized by the same energy but with different flux, charge
and angular momentum.

\section {Nonrelativistic theory}

\subsection { Nonrelativistic Limit }

 In this section, we want to study the nonrelativistic limit of
our model. We consider $\bar{h}=1$ and keep the velocity of light c
explicitly since we are interested in the nonrelativistic limit
$c \rightarrow \infty$. All other conventions we follow in this
section are same as described in section II. A. We want to compare
our model
directly with that of Jackiw and Pi\cite{pi}. So, for convenience,
we remove the factor ${1 \over 2}$ multiplying with the scalar
kinetic energy term
in the Lagrangian (\ref{eq:rl}). The subsequent effect of this
change on
the dielectric function as well as on the scalar potential have been
taken into account. After kepping
the velocity of light $c$ explicitly and making the above mentioned
change in the Lagrangian (\ref{eq:rl}), we have
\bea
{\cal L} \ & = & \ - \ {\ 1 \over 4} G({\mid \phi \mid})
F_{\mu \nu}  F^{\mu \nu}
+ \ D_{\mu} {\phi} (D^{\mu}{\phi})^* \ + \ {\kappa \over 4}
\epsilon^{\mu \nu \lambda}  A_{\mu}  F_{\nu \lambda}\nonumber \\
& & - \ {{\ 4 e^4 v^{4-4 \gamma}} \over {\ (\gamma+1)^2 C_{0}^2
\kappa^2
c^4}} {\mid \phi \mid}^2 (v^2 - {\mid \phi \mid}^2)^{\ 2 \gamma}
\label{eq:nrl}
\eea
\noindent where the covariant derivative is given by
\be
 D_\mu \phi \ = \ (\partial_{\mu}  \ - \ {\ i e \over c} A_\mu
 - \ {{\ i g} \over {\ 4 c}} G(\mid \phi \mid)
\epsilon_{\mu \nu \lambda} F^{\nu \lambda})\phi
\label{eq:cd2}
\ee
\noindent Nontopological vortex solution exists in this theory for any
positive odd integer $\gamma$, when the dielectic function
$G(\mid \phi \mid)$ assumes the following form
\be
G(\mid \phi \mid) \ = \ {{\kappa^2 c^2} \over
{2 e^2 {\mid \phi \mid}^2}}
-{{\kappa^2 c^2 C_{0} v^{2 (\gamma-1)}} \over {2 e^2
{\mid \phi \mid}^2 (v^2-{\mid \phi \mid}^2)^{\gamma-1}}}
\label{eq:df1}
\ee
\noindent and $C_{0} > 0$.

Recall the fact that the mass of a scalar field is defined through the
coefficient of the quadratic term in a scalar field potential, which
is $m^2 c^2$ in this case. Comparing
with the potential term in (\ref{eq:nrl}), we find that $v^2$ should
have the value
\be
v^2 \ = \ {{\ m c^3 {\mid \kappa \mid} C_{0} (\gamma+1)}
\over {\ 2 e^2}}
\label{eq:vv}
\ee
\noindent Now the matter part of the
Lagrangian density (\ref{eq:nrl}) can be written in terms
of the scalar
field mass $m$
\bea
{\cal L}_{matter} \ & = & \ {\ 1 \over c^2} {\mid {\partial_t-
i e A_{0}-
{{\ i g} \over 2} F_{12} G({\mid \phi \mid})} \mid}^2
- \ ({ D_{i} \phi}) ({ D_{i} \phi})^*\nonumber \\
& & - \ m^2 c^2 {\mid \phi \mid}^2 \
+ \ {{4 \gamma e^2 m } \over {\ (\gamma+1) c C_{0}
{\mid \kappa \mid}}} {\mid \phi \mid}^4\nonumber \\
& & + \ {{4 e^4} \over {(\gamma+1)^2 C_{0}^2 \kappa^2
c^4}} \sum_{p=2}^{2 \gamma} (-1)^p \ ^{2 \gamma}C_{p} \ v^{2 (2-p)}
{\mid \phi \mid}^{2 (p+1)}
\label{eq:nrl1}
\eea
\noindent To consider the nonrelativistic limit,
we first substitute in
(\ref{eq:df1}) and (\ref{eq:nrl1})
\be
\phi \ = \ {\ 1 \over {\sqrt{2 m}}} [e^{-i m c^2 t} \psi \ +
 \ e^{i m c^2 t} \tilde{\psi}^*]
\label{eq:phi}
\ee
\noindent and drop all terms which oscillate
as $c \rightarrow \infty$. Keeping only dominant inverse powers
of $c$ and
setting $\tilde{\psi}=0$ in order to work in the zero antiparicle
sector ( since particles and antiparticles are separately
conserved ), we obtain the nonrelativistic dielectric function
\be
G(\rho) \ = \ {{\kappa^2 c^2 m (1-C_{0})} \over {e^2 \rho}}
\label{eq:df2}
\ee
\noindent as well as the Lagrangian density
corresponding to the relativistic Lagrangian (\ref{eq:nrl}),
\bea
{\cal L} \ & = &  \ \psi^* i (\partial_t-i e A_{0} -
{{\ i g} \over 2} \ {{\kappa^2 c^2 m (1-C_{0})} \over {
\rho e^2}} \ F_{12} ) \psi \ -
{\ 1 \over {2 m}} (D_i \psi) (D_i \psi)^*\nonumber \\
& & +{{\gamma e^2  \rho^2} \over {\ (\gamma+1) m c C_{0}
{\mid\kappa\mid}}}
- \ {{\kappa^2 c^2 m (1- c_{0})} \over { 4 e^2 \rho}} F_{\mu \nu}
F^{\mu \nu} \ + {\kappa \over 4} \epsilon_{\mu \nu \lambda}
A^\mu F^{\nu
 \lambda}
\label{eq:nrl3}
\eea
\noindent where $\rho=\psi^* \psi$ is the particle density.
The Lagrangian (\ref{eq:nrl3}) describes the second
quantized
version of a fixed number of nonrelativistic particles moving
in $\delta$-function potential and interacting with massive
relativistic photons. Note that the effect of the scalar potential
with higher values of $\gamma$
is to change the strength of the $\delta$-function interaction
in the nonrelativistic limit.
Since $C_{0}$ is positive in order to have nontopological vortex
solutions in the relativistic theory and $m$ is defined to
be positive through (\ref{eq:vv}), the $\delta$-function interaction
is attractive.

 We obtain the equation of motion by varying the Lagrangian
(\ref{eq:nrl3}) with respect to $\psi^*$ and $A_\mu$ respectively
\bea
- i (\partial_t - i e A_{0} -
{{\ i g} \over 2} \ {{\kappa^2 c^2 m (1-C_{0})} \over {e^2 \rho}}
F_{12}) \psi \ & = & \ {1 \over 4} {{\kappa^2 c^2 m (1-C_{0})}
\over {e^2 \rho^2}} [ \ F_{\mu \nu} F^{\mu \nu}\nonumber \\
& & + {\ g \over { e c}}
\epsilon_{\mu \nu \alpha} J^\mu F^{\nu \alpha} \ ] \psi\nonumber \\
& & + {\ 1 \over {2 m}} D_i D_i \psi \ +
{{2 \gamma e^2 \rho} \over {(\gamma+1) m c C_{0} {\mid \kappa \mid}}}
\psi
\label{eq:nem1}
\eea
\be
\epsilon_{\mu \nu \lambda} \ {\partial^\mu} \ [
\ {{\kappa^2 c^2 m (1-C_{0})} \over {e^2 \rho}} \ (F^\lambda
\ + {\ g \over 2 e c} J^\lambda)] \ = {\ 1 \over c} \ J_\nu \ -
\ \kappa F_\nu
\label{eq:nem2}
\ee
\noindent where $J_\nu$ is the current density
\bea
J_\nu \ & = & \ (-e c \rho, J_i)\nonumber \\
& = & [- e c \psi^* \psi, \ -{{i e} \over {2 m}} \{\psi^* D_i \psi-
\psi (D_i \psi)^*\}]
\label{eq:ncd}
\eea
\noindent The energy is given by
\bea
E \ & = & \ c {\cal \int} d^2 x [{1 \over {2 m}} (\bigtriangledown_i
\psi)
(\bigtriangledown_i \psi)^* \ - \ {{ \gamma e^2 \rho^2} \over
{(\gamma+1) m c C_{0}
{\mid \kappa \mid}}}\nonumber \\
& & + {{\kappa^2 c^2 m (1-C_{0})} \over {2 e^2 \rho}}
(F_{12})^2 \
+ {{\kappa^2 c^2 m C_{0} (1-C_{0})} \over {2 e^2 \rho}}
(F_{0 i})^2 \ ]
\label{eq:nener}
\eea
\noindent where $\bigtriangledown_i=\partial_i-{{i e} \over c} A_i$
includes only the gauge potential. Like relativistic case notice
that the solution to the equation
\be
J_\nu = \kappa c F_\nu
\label{eq:nrfgf}
\ee
also solves gauge field equation (\ref{eq:nem2}) provided
$g=-{{2e} \over \kappa}$. The time component
of equation (\ref{eq:nrfgf}), i.e,
the ${\bf CS}$ modified Gauss law gives
$B \ = \ {e \over \kappa} \rho$, where $B$ is the magnetic field.
The immediate consequence of equation
(\ref{eq:nrfgf}) is that any solution with flux ${\Phi}$ also
carries charge $Q = - \kappa c \Phi$.
The space component of equation (\ref{eq:nrfgf})
can be written in component form as
\bea
E^i \ & = & \ - \partial_i A^{0} -
{1 \over c} \partial_t A^i\nonumber \\
& & = {1 \over {c \kappa}} \ \epsilon^{i j} \ J^j
\label{eq:nelec}
\eea
\noindent This equation can be further simplified as,
$E^i = {1 \over {c \kappa}} \epsilon^{i j} J^j  =
{1 \over {C_{0} c \kappa}} \epsilon^{i j} {\tilde{J}}^j$.
Note that the expression for the electric field components is
exactly same as that of nonrelativistic pure
{\bf CS} case\cite{pi} except the factor ${1/C_0}$. We rewrite
equation (\ref{eq:nem1}) as
\bea
- \ i (\partial_t - i e A_{0} -
{{\ i g} \over 2} \ {{\kappa^2 c^2 m (1-C_{0})} \over {e^2 \rho}}
F_{12}) \psi \  & = &  \
{\ 1 \over {2 m}} D_i D_i \psi\nonumber \\
& & - \ {\ 1 \over 4}
{{\kappa^2 c^2 m (1-C_{0})} \over {e^2 \rho^2}}
F_{\mu \nu} F^{\mu \nu} \psi\nonumber \\
& & + {{2 \gamma e^2 \rho} \over {(\gamma+1) m c C_{0}
{\mid \kappa \mid}}} \psi
\label{eq:nem3}
\eea

\subsection {Self-dual Solution}

\noindent Now we seek rotationally symmetric static soliton
solutions in the
system described by the equations (\ref{eq:nrfgf})
and (\ref{eq:nem3}). We choose the following Ansatz
\be
\psi \ = \ e^{-i n \theta} \rho^{1 \over 2} (r) \ ,
\vec{A}= \ - \ \hat{\theta} {{c (a(r) - n )} \over {e r}} ;
\ A_{0}= h(r).
\label{eq:nansatz}
\ee
\noindent Substituting (\ref{eq:nansatz}) into the equation
(\ref{eq:nelec}), and (\ref{eq:nem3}), we have
\be
h^\prime(r)= \ - {{e \ a \ \rho} \over {\kappa \ m \ c \ r \ C_{0}}}
\label{eq:nem4}
\ee
\be
{1 \over r} {\partial \over {\partial r}}(r {{\partial}
\over {\partial r}}
\log \rho) \ = - \
{{8 \gamma e^2 \rho} \over {(\gamma+1) c C_{0} {\mid \kappa \mid}}}
- \ 4 m e A_{0} -2 m^2 c^2 (1-C_{0}) + \
{{2 a^2} \over {r^2 C_{0}}} - {{\rho^\prime}^2 \over
{2 \rho^2}}
\label{eq:nem5}
\ee
\noindent where use have been made of the relation $B = {e \over
\kappa} \rho$. The energy functional can be written as,
\be
E \ = c {\cal \int} d^2 x \ [ \ {1 \over {8 m}}
({\rho^\prime \over \sqrt{\rho}})^2
+ \ {{a^2 \rho} \over {2 m C_{0} r^2}} -
{{\gamma e^2  \rho^2} \over {(\gamma+1) m c C_{0} {\mid \kappa \mid}}}
+ \ {{ m c^2 (1-C_{0}) \rho} \over 2} \ ]
\label{eq:nem6}
\ee
\noindent Using Bogomol'nyi trick we rewrite $E$
\bea
E \ & = & \ c {\cal \int} d^2 x \ [ \ {1 \over {2 m}}
({\rho^\prime \over {2 \sqrt{\rho}}} \pm
{{ a \ \sqrt{\rho}} \over {\sqrt{C_{0}} \ r}})^2
\ \mp \ {{e^2 \rho^2} \over {2 m c \sqrt{C_{0}}
 \kappa}}\nonumber \\
& & - {{\gamma e^2  \rho^2 } \over {(\gamma+1) m c C_{0}
{\mid\kappa\mid}}} \ ]
+ \ {{\kappa m c^3 (1- C_{0})} \over {2 e}} \Phi
\label{eq:nem7}
\eea
\noindent There is a lower bound on the energy
$E \geq {{\kappa m c^3 (1-C_{0})} \over {2 e}} \ \Phi$ provided we
choose
\be
\sqrt{C_{0}} = {{2 \gamma} \over {(\gamma+1)}}
\label{eq:cons}
\ee
\noindent  and take the upper sign
in case $\kappa$ negative while the lower sign for $\kappa$ positive.
The bound is saturated when the following first order
equation holds
\be
{\rho^\prime \over {2 \sqrt{\rho}}} \ = \ \mp {{ a \ \sqrt{\rho}}
 \over {\sqrt{C_{0}} \ r}}
\label{eq:nbe}
\ee
\noindent Notice from equations (\ref{eq:nem7}) and (\ref{eq:cons})
that when $\gamma = 1$, and hence ${C_{0}=1}$ (i.e, for pure
$\bf {CS}$ theory in the NR limit), static self-dual soliton solution
exists with zero energy. However the effect of the
Maxwell term in the ${\bf NR}$ limit can be seen for $\gamma > 1$.
As is clear from equations (\ref{eq:nem7}) and (\ref{eq:cons}) the
lower bound on the energy is saturated at nonzero finite value.
Since $C_{0} > 1$ for $\gamma > 1$
and the positive flux corresponds to the lower sign in
(\ref{eq:nem7}) and vice versa; the minimum energy is always negative.

 The
decoupled equation for the charge density $\rho$ is,
\be
\bigtriangledown^2 \log \rho(r) \ = \ \pm {{2 \tau}
\over \sqrt{C_{0}}} \ \rho(r)
\label{eq:dme}
\ee
\noindent where $\tau={e^2 \over {c \kappa}}$.
The equation (\ref{eq:dme}) is the Liouville equation which is
completely integrable. It is worth while to mention that for the whole
class of the dielectric function for which the nontopological vortex
solutions
exists in the relativistic theory, the charge density solves the same
Liouville equation (\ref{eq:dme}) in the nonrelativistic limit.
However, both the
soliton energy and the Liouville equation are parametrized by $C_{0}$
which is determined in terms of $\gamma$.
The Liouville equation admits nonsingular nonnegative solutions
for $\rho(r)$ when the numerical constant on the right hand side of
(\ref{eq:dme}) is negative.
For both the self-dual
and anti self-dual solutions, the numerical constant
$\pm {{2 \tau} \over \sqrt{C_{0}}}$ is indeed
negative according to the sign convention we fixed after the equation
(\ref{eq:nem7}). The matter density $\rho(r)$ that solves
(\ref{eq:dme}) is
\be
\rho(r) = {{4 (n+\sqrt{C_{0}})^2 }
\over {\sqrt{C_{0}} \ \tau \ r^2}} \
[ \ ({r_{0} \over r})^{{n \over \sqrt{C_{0}}} +1}
+ ({r \over r_{0}})^{{n \over \sqrt{C_{0}}} +1} \ ]^{-2}
\label{eq:liv}
\ee
\noindent where $r_{0}$ is a parameter describing the solution.
After substituting (\ref{eq:liv}) into equation (\ref{eq:nbe}),
we find
\be
a(r) = \pm \ [ \ \sqrt{C_{0}} - \ (n +\sqrt{C_{0}}) \
{{ ({r_{0} \over r})^{{n \over \sqrt{C_{0}}} +1}
- ({r \over r_{0}})^{{n \over \sqrt{C_{0}}} +1}} \over
{({r_{0} \over r})^{{n \over \sqrt{C_{0}}} +1}
+ ({r \over r_{0}})^{{n \over \sqrt{C_{0}}} +1}}} \ ]
\label{eq:liv1}
\ee
\noindent Note that both $f(r)$ and $a(r)$ depends on $C_{0}$
nontrivially. The charge density $\rho(r)$ vanishes both at the
origin and at the assymtotic infinity for any $C_{0}$ satisfying
the relation (\ref{eq:cons}). However, for fixed $r_{0}$, $n$ and
$\tau$, the rate of fall off at both the limit is
slower for higher values of $C_{0}$, i.e, for the
higher values of $\gamma$. As
$r \rightarrow 0$, $a(r) \rightarrow \mp n$ ensuring the
nonsingularity of the gauge field $A_{2}$. At large distances
$a(r)$ approaches the value $ \pm \ (n +
2 \sqrt{C_{0}})$ which, interestingly enough, is exactly
the lower bound on $\alpha$ in the relativistic theory as shown
in subsec. II. D. In other words, $\alpha$ saturates its lower
bound in the nonrelativistic limit. These soliton solutions are
characterized by
the magnetic flux $\Phi= \mp {{4 \pi c}\over e} \ (n+\sqrt{C_{0}})$
, the charge $Q=-\kappa c \Phi$ and the angular momentum
$J \ = \ \mp {{2 \kappa c} \over e} \ \sqrt{C_{0}} \ \Phi$.
It is obvious from equation (\ref{eq:cons}) that $C_{0} < 4$
for any value of $\gamma$. So, the magnetic flux, the charge
, the angular momentum and the soliton energy are finite for
any $\gamma$ and in
particular, $(n+1) \ \leq \ {e \over {4 \pi c}} \ \mid\Phi\mid \ <
\ (n+2)$.

To show that the first order Bogomol'nyi equations are consistent
with the second order equation (\ref{eq:nem5}), note that the
current density $J_2$ can be written in London form,
\be
J_{2} \ = \ {1 \over C_{0}} \tilde{J_{2}} \ = \ -
{e \over {m C_{0} r}} \rho(r) a(r)
\label{eq:london}
\ee
\noindent Since we are considering only static soliton solution, we
choose $J^2$ to be transverse. Following Jackiw and Pi\cite{pi},
$A_0$ can be fixed as,
\be
A_0 \ = \ - {{m c^2 (1- C_{0})} \over {2 e}} \ \pm
{{e \rho} \over {2 m c \sqrt{C_{0}} \kappa}}
\label{eq:nem8}
\ee
\noindent Now it is easy to see that the first order equations are
consistent
with the second order equation (\ref{eq:nem5}).

\section{Conclusion}

 We have considered a generalization of the abelian Higgs model with
a ${\bf CS}$ term by multiplying a dielectric function with the
Maxwell term and taking generalized covariant derivative. We have
obtained Bogomol'nyi bound for this model, when certain relation
among the coupling constant holds and the dielectric function assumes
a particular form. The Bogomol'nyi equations of pure ${\bf CS}$
vortices and of Torres model are obtained as two special cases, up to
a scale transformation of the variables. We have also obtained a novel
new type of topological vortex solutions for which flux, charge and
angular momentum need not be
quantized, even though the energy is quantized.
These toplogical vortex solutions are infinitely degenerate in each
sector and differ from each other by flux, charge and angular
momentum. We have been sucessfull to put both  upper and
lower bound on the magnetic flux, the charge and the angular momentum
of these topological vortices by deriving
sum rules. We have also studied nontopological vortex solutions in
this model along with their physical properties.

 Furthermore, we have
considered the ${\bf NR}$ limit of our model and obtained static
self-dual soliton solutions. Though both the ${\bf NR}$ pure
${\bf CS}$ and ${\bf MCS}$ self-dual soliton solutions are zero
energy
configuration, the self-dual soliton solutions in {\bf GMCS} model
saturates the lower bound at some nonzero finite value of the energy.
For the whole class of scalar
potential for which nontopological vortex solution exists in the
relativistic ${\bf GMCS}$ theory, the charge density in
the ${\bf NR}$ theory satisfies
the same Liouville equation. Since the contribution of the
scalar potential
with higher values of $\gamma$ to the ${\bf NR}$ theory is to change
the strength
of the $\delta$-function interaction, both the Liouville equation
and the soliton energy are parameterized by $C_{0}$ which is
dependent on
$\gamma$. In fact, the relation between
$C_{0}$ and $\gamma$ which is not present
in the relativistic theory comes as a constraint in order
to have self-dual solution in the nonrelativistic limit.

 For further investigations, this model raises a number of interesting
questions. First of all, it is well known that for the Lagrangian
considered by Torres (i.e, G=1), the physical photon mass does not
receive one-loop radiative correction at the Bogomol'nyi Limit
\ $ (i.e, g=-{{2 e} \over \kappa})$\cite{wallet}.
Does the physicsl photon mass in ${\bf GMCS}$ model receives one
loop radiative correction at the Bogomol'nyi limit,
when only renormalizable potential are allowed?
The bogomol'nyi equations are often manifestation of $N=2$
superymmetry\cite{susy}.
Is it true for {\bf GMCS} vortices also?
If the radiative correction to photon mass vanishes at one loop level
at the Bogomol'nyi limit
and a $N=2$ superfield formalism is possible for {\bf GMCS} model,
could one correlate these two phenomena?

 The Bogomol'nyi equations we obtained are quite
different in nature from the corresponding equations for vortices in
other well known models. Naturally it is worthwhile to study whether
the usual technique for showing the uniqueness and existence of
soliton solutions goes through in this case or not. Also it
would be interesting to know the total number of independent zero
modes present in this model
for generic dielectric function and the corresponding scalar
potential. Above all, the most interesting thing it would be if
this model can be realized in any planar condesed matter system where
dielectric function plays major role.

\acknowledgements
I thank Prof. Avinash Khare for valuable discussions and critically
going through the manuscript. I thank Munshi Golam Mustafa for helping
me in numerical computation.

\begin{figure}

\caption{ A plot of $f(r)$ (solid line) and $a(r)$ (dashed line) for
$\gamma=3$ nontopological vortices with $n=1$, $\alpha= 4.27$ .}

\caption{ A plot of $f(r)$ (solid line) and $a(r)$ (dashed line) for
$\gamma=3$ nontopological vortices with $n=2$, $\alpha= 5.51$.}

\caption{ A plot of $f(r)$ (solid line) and $a(r)$ (dashed line) for
$\gamma =3$ nontopological soliton with $\alpha= 4.22$ and $n=0$.}

\caption{ A plot of $f(r)$ (solid line) and $a(r)$ (dashed line)
for $\gamma=3$ topological vortices with (I) $n=1, \beta=0.78$;
(II) $n=2, \beta=1.91$; (III) $n=2, \beta=1.53$ and (IV) $n=2,
\beta=1.23$ .}

\end{figure}

\end{document}